\def\mincir{\raise -2.truept\hbox{\rlap{\hbox{$\sim$}}\raise5.truept
\hbox{$<$}\ }}
\def\magcir{\raise -2.truept\hbox{\rlap{\hbox{$\sim$}}\raise5.truept
\hbox{$>$}\ }}

\documentclass[12pt,preprint]{aastex}

\shorttitle{Continuum Variability of AGNs}
\shortauthors{Tr\`evese \& Vagnetti}

\slugcomment{To appear in Ap. J., January 2002}

\begin{document}

\title{Quasar Spectral Slope Variability in the Optical Band}

\author{Dario Tr\`evese}
\affil{Dipartimento di Fisica, Universit\`a di Roma ``La Sapienza'',\\
Piazzale A. Moro 2, I-00185 Roma, Italy, dario.trevese@roma1.infn.it, and\\
Dipartimento di Fisica, Universit\`a di Roma ``Tor Vergata'',\\
Via della Ricerca Scientifica 1, I-00133 Roma, Italy}

\and

\author{Fausto Vagnetti}
\affil{Dipartimento di Fisica, Universit\`a di Roma ``Tor Vergata'',\\
Via della Ricerca Scientifica 1, I-00133 Roma, Italy,
fausto.vagnetti@roma2.infn.it}

\begin{abstract}
We performed a new analysis of $B$ and $R$  light curves of a sample of PG 
quasars. We confirm the variability-redshift correlation and its explanation
in terms of spectral variability, coupled with the increase of rest-frame 
observing frequency for quasars at high redshift.
The analysis of the instantaneous spectral slope for the whole quasar 
samples indicates both an inter-QSO and intra-QSO variability-luminosity
correlation. Numerical simulations show that the latter correlation
cannot be entirely due to the addition of the host galaxy emission to a nuclear
spectrum of variable luminosity but constant shape, implying a spectral 
variability of the nuclear component. Changes of accretion rate are
also insufficient to explain the amount of spectral variation, while
hot spots possibly caused by local disk instabilities can explain 
the observations.

\end{abstract}


\section{INTRODUCTION}

Although variability plays a key role in constraining the size of the central
engine of active galactic nuclei, 
yet its  physical origin remains substantially unknown.
Even restricting to the class of non-Blazar objects,
the most diverse mechanism have been proposed in recent years, including
gravitational lensing due to intervening matter \citep{haw96},
supernovae explosions \citep{are97}, instabilities in the accretion 
disk \citep{kaw98}  and star collisions \citep{tor00}.
For a small number of low redshift objects, a multi-wavelength monitoring with 
adequate time sampling and resolution allows the interpretation of changes 
of the spectral energy distribution (SED) in terms of an interplay of emission
components with different spectral and variability properties
(see \cite{umu97} 
for a general review and \cite{cou98} for the case of 3C273).
In the near infrared-optical-UV bands, variability studies indicate a 
hardening of the spectrum in the bright phase
\citep{cut85,ede90,kin91,pal94}.
So far, however, most of the statistical information on AGN variability, 
derives from single-band light curves 
of magnitude limited samples of objects
\citep{ang72,bon79,haw83,tre89,cri90,tre94,hoo94,ber98}. 
In the case of magnitude limited samples the analysis is 
complicated by the strong luminosity-redshift (L-z) correlation, caused by
the crowding of objects towards the limiting flux.
As a consequence it is difficult to isolate any intrinsic 
variability-luminosity (v-L) and variability-redshift (v-z) correlation.
Moreover, the results of these analyses  depend on the specific variability
index adopted, as shown by \cite{gia91} who found a positive v-z correlation 
through a variability index defined on the basis of the rest-frame structure 
function. The existence of an average increase of variability with redshift 
was later confirmed by \cite{cri96}. It turns out to be consistent with the 
suggestion of \cite{gia91} that QSOs at high redshift appear more variable
since they are observed at a higher rest-frame frequency, where the 
variability is stronger (hardening in the bright phase and vice versa). 
A direct statistical evidence  of the spectral variability,
in terms of an average change of the B-R color with 
the variation of the  B magnitude, was found by \cite{giv99} as part of a 
statistical analysis of the B and R light curves of a sample of PG QSOs.
A similar evidence has been found
by \cite{tre01} for the faint QSO sample of SA 57 \citep{kkc}.
Unfortunately, most (if not all) of the variability mechanisms proposed so 
far imply a hardening of optical-UV  spectrum in the bright phase, so that 
they cannot be  discriminated on  purely qualitative grounds.
As a first step towards constraining variability  models quantitatively
\cite{tre01} compared with the observations a simple model where both 
spectral slope changes and
brightness variations are due to temperature changes of an emitting black body.
In the present paper, we present a new analysis of the B and R light curves
of a sample of PG QSOs made available by \cite{giv99}, 
we analyze the consistency with previous results, and consider possible different 
sources of spectral variability.
The paper is organized as follows: in section 2 we summarize the 
characteristics of the \cite{giv99} data, in section 3 we analyze the 
v-z correlation, in section 4 we discuss the parameters adopted for the 
spectral variation analysis, in section 5 we consider the effect of the 
host-galaxy on SED variations, in section 6 we compare the observed spectral 
variations with those produced by changes of the accretion rate,
and discuss the consistency
of the observations with a simple model consisting of  hot spots
on the accretion disk. Section 7 contains a summary and 
the conclusions.
We adopt $H_o= 50 {\rm ~km~s^{-1}~Mpc^{-1},~q_o=0.5}$, unless otherwise stated.

\section{THE DATA}

The present analysis is based on the light curves made available
to the community by the Wise Observatory group \citep{giv99}.
The sample consist of 42 PG QSOs selected to be 
nearby, i.e. $z < 0.4$, and bright, i.e. $B < 16$ mag.

The observations were made in the Johnson-Cousins
B and R bands with the 1 m Wise Observatory telescope. The total duration of 
the campaign was 7 years and the median observing interval of the objects
was 39 days.
The r.m.s. photometric uncertainty is $\sim 0.01$ mag and 
$\sim 0.02$ mag in the B and R bands respectively. 
We refer to the original paper of \cite{giv99} for all  
details concerning  observations,  calibration etc.
The full B and R light curves of the entire sample were made 
available in electronic form.

\cite{giv99} present the analysis of the correlation of different 
variability properties with other properties like luminosity, redshift,
radio power, various line intensities and the X-ray spectral slope.
In particular they show a correlation  between the color changes
$\Delta (B-R)$ and the brightness variations $\Delta B$ and $\Delta R$,
corresponding to an average hardening of the spectrum in the bright phase
(and vice versa). They do not find a  correlation between variability and 
redshift, at variance with \cite{gia91}, \cite{tre94}, \cite{cri96}, but they
ascribe this to the difficulty of disentangling v-z, v-L and L-z
correlations in the sample. We stress that this is made particularly
difficult by the small redshift interval spanned by the sample.
We perform a new analysis of both the v-z correlation and the spectral slope
changes in the following sections. For these purposes we have dereddened the
data, using the extinction calculator of the NASA Extragalactic Database (NED).

\section{VARIABILITY-REDSHIFT CORRELATION}

To measure the amplitude of variability we define, for each object,
the first order structure function, as in \cite{dic96} 
(but omitting the subscript ``1''):

\begin{equation}
S(\tau,\Delta \tau)=[({\pi\over 2}\overline{|m(t+\tau)-m(t)|}^2
-\sigma_n^2)]^{1\over 2}
\end{equation}

\noindent
where $m(t)$ is either the $B$ or the $R$ magnitude, $t$ is the
rest-frame time, $\tau$ is the time lag between the observations,
$\sigma_n$ is the relevant r.m.s. noise and the bar indicates
the average  taken
over all the pairs of observations lying in the time interval
$\tau \pm \Delta \tau$.
In this definition we adopt the square average of the absolute values of the
difference instead of the average of the square difference as in 
\cite{dic96}, since the former 
quantity is less sensitive to outliers \citep{hoo94}. The $\pi/2$ factor
normalizes $S$ to the r.m.s. value in the case of a Gaussian distribution.
The adopted value of $\Delta \tau$ is the result of a trade-off between
time resolution and statistical uncertainty.

In the following we define four variability 
indices $S_i(\tau \pm \Delta \tau)$, with
$i=B,R$, and $\tau=0.3  \pm 0.09$ yr, $2.0  \pm 0.6$ yr. The subscripts
$B$ and $R$ refer to the observing band and
the values of $\tau$ and $\Delta \tau$ have
been chosen for comparison with previous analyses \citep{dic96}.
None of these four indices shows a significant correlation with redshift
when the whole sample is considered, confirming the result of \cite{giv99}.

However, to disentangle the v-L, v-z and L-z correlations, we can restrict
the analysis to a magnitude bin $-23.5 < M_B  < -22.5$, around the average
absolute magnitude of the sample $<M_B>=-22.75$. The result is shown in
Figure 1. In this case we find a v-z correlation coefficient $r_{v,z}=0.39$,
which is marginally 
significant ($P(>r)=0.09$) despite the small number of objects (19) in the bin.
To examine the dependence of variability on redshift
we take the ensemble averages of the four variability indices defined above, 
over the same subsample of 19 objects. For each observing band we compute the
average rest-frame observing frequency of the sample.
To compare the result with \cite{dic96} we must take into account
the dependence of variability on magnitude.
For this purpose, we reduce $S$ by an amount 
$\Delta S = (\partial S / \partial M_B)~\Delta M_B$
where for $S(M_B)$ we adopt model A of \cite{cri96} and $\Delta M_B$
is the difference between the average absolute magnitudes of the present sample
and the sample of \cite{dic96}. 
Figure 2 shows the increase of variability
with the observing frequency. At each frequency variability is larger
for larger time lag $\tau$ (due to the increase of the structure function
from $\tau=0.3$ yr
to $\tau=2$ yr). The new points, obtained from the present analysis,
are consistent with the previous results, which were  obtained from different 
samples and observing frequencies. The general trend can be quantified as: 

\begin{equation}
\partial S_i / \partial \log \nu_{rest} = \partial
S_i / \partial \log (1+z) \simeq 0.25 - 0.3
\end{equation}

\noindent
and is consistent with \cite{gia91},  \cite{cri96} and \cite{dic96}.
Therefore the increase of variability with redshift and its interpretation in
terms of an average increase of the amplitude of variability with frequency are confirmed
by the present results despite their poor statistical basis.

\cite{haw96} finds that the variability timescale is independent
of redshift, i.e. it is not affected by cosmological time dilation. He suggests
that the effect can be explained in the framework of
gravitational microlensing caused by intervening matter.
\cite{haw01} estimates that a possible decrease of variability
{\it timescale} with frequency is not sufficient
to compensate for the cosmological time dilation.
The frequency dependence of variability {\it amplitude}
can also compensate, at least in part, for time dilation.
We note that from Eq. 2 we can estimate a vertical shift of $\sim 0.1$
between $B$ and $R$ power spectra, approximately consistent with
the horizontal shift of $\sim 0.06$ found by 
\cite{haw01} (see his Figure 3).
However, a detailed evaluation of the 
amplitude shift of the variability power spectrum as a function of redshift would require 
the knowledge of the redshift, luminosity, and observing time distributions of the sample.

We also stress that the frequency dependence of microlensing
\citep{ale95} should be compared with the results shown  in Fig. 2 and Eq. 2.

\section{THE SPECTRAL SLOPE $\alpha$ AND THE SPECTRAL VARIABILITY PARAMETER $\beta$}

The average increase of variability with frequency can be interpreted in 
terms of a change of the spectral slope with luminosity. However the 
statistical evidence discussed in the previous section, and in
\cite{gia91}, \cite{dic96}, is only indirect and must be confirmed by observations 
of individual objects in at least two bands. This was done by \cite{giv99}
in terms of correlation of color changes $\Delta(B-R)$ with brightness
variations $\Delta B$ or $\Delta R$ and by \cite{tre01} in terms of changes 
of the spectral slope $\alpha$, defined by $L_{\nu} \sim \nu^{\alpha}$, 
where $L_{\nu}$ is the intrinsic power per unit frequency. The latter work
is based on $U$,$B_J$,$F$, $N$ observations at two epochs of the sample of QSOs
of the Selected Area 57. The increase of spectral slope is computed as an 
average value over the ensemble of objects, and compared with a simple model 
consisting of a black body subject to small temperature changes.
The data of \cite{giv99}, which contain on average 40 $B$ and $R$ observations
of each QSO, allow the statistical analysis of spectral slope changes of each
object, for which we compute the instantaneous slope:

\begin{equation}
\alpha(t) \equiv \log (L_{\nu_B}/L_{\nu_R})/\log(\nu_B/\nu_R)=
-\frac{0.4[(B-R)-(B_o-R_o)]} {\log \frac {\lambda_R}{\lambda_B}} - 2,
\end{equation}

\noindent where $m_o=-2.5 \log f_o$ are the zero points of B and R photometric bands 
respectively \citep{cox00}.
Taking into account that typical variability time scales are $\approx 1$ yr
(see e.g. \cite{tre94}) we regard as simultaneous the observations within a 
time interval of
9 hours. In Figure 3 we report the instantaneous value of the spectral slope 
$\alpha$ as a function of the intrinsic luminosity $L_{\nu_B}$. Each small 
cloud of points represents a single QSO in different luminosity states.
The relevant regression lines are reported on each cloud. They show, with a 
few exceptions, a positive correlation between the spectral slope and the 
intrinsic luminosity, which we call intra-QSO $\alpha-L$ correlation.
The distribution of clouds in Figure 3 also shows that the average slope of
each QSO tends to be larger for brighter objects,
forming a sort of QSO main sequence in the $\alpha-L$ plane \citep{tre01a}. 
This inter-QSO $\alpha-L$ correlation
can be quantified considering for each QSO the average values
$\langle\alpha\rangle$ and $\langle L_{\nu_B}\rangle$
of the slope and luminosity. The correlation
$r_{\alpha-L}=0.58$ is highly significant: $P(>r)=6 \cdot 10^{-5}$.
 
Figure 3 might suggest that both
correlations are produced by the same physical mechanism, e.g. an increase of 
the temperature of the emitting gas (\cite{tre01}, see also \cite{pal94}).
However, in order to try any comparison with possible models, 
a precise quantification of the above effects is needed.
In fact, a variety of mechanisms might in principle be responsible for
a hardening of the spectrum in the bright phases.
For instance, if AGNs are powered by supernovae explosions and variability
is caused by a ``Christmas tree'' effect, then there will be an excess of blue 
emission in the bright phase \citep{are97,cid00}.
Even gravitational lensing \citep{haw96}, usually thought of as achromatic, 
can produce a stronger variability in the blue than in the red band 
since the amplification depends on the size of the accretion disk 
which is larger in the red than in the blue band \citep{ale95}.

Also hot spots due to instabilities of the accretion disk \citep{kaw98} are 
likely to enhance the blue emission.
Finally the spectral energy distribution of the host galaxy, 
which is redder than the AGN and gives a
stronger contribution in the faint AGN phase, can produce a similar effect.

In order to quantify the spectral variations we define a spectral variability
parameter (SVP) representing the spectral slope changes per unit 
log-luminosity change:
\begin{equation}
\beta(\tau)\equiv\frac{\alpha(t+\tau)-\alpha(t)}{\log L_B(t+\tau)-\log L_B(t)},
\end{equation}
where $L_B(t)$ is the luminosity in the $B$ band and $\tau$ is an appropriate
time delay. In fact we expect that different physical phenomena are causing 
different SED changes
on the relevant time scales, which go at least from days to more than ten years. 
With the available light curves we have computed, for each QSO, 
$\beta(\tau_{ij})$
with $\tau_{ij}\equiv t_i-t_j$ ,$i,j=1,N$ representing all the possible time 
differences between the N points of the light curve. Figure 4 represents 
$\beta(\tau_{ij})$ for two of the 42 QSOs considered, PG0804+762 and
PG1354+213, taken as examples of good 
and poor time sampling respectively. For each bin, the mean value of the SVP is also shown.
The uncertainty is the r.m.s. variation of the mean.
No obvious trend is seen looking at similar plots for the 
entire sample: the average $\beta$ values stay almost constant, within the 
uncertainty, at least for $\tau \mincir 1000$ d. 
Clearly, bins at large $\tau$ are less populated. 
An increase or decrease of $\beta$ appears for 
some objects at larger $\tau$ where, however, the statistic becomes poor.
Apparently, an analysis of possible systematic $\beta$ changes on time scales 
$\tau \magcir 3$ yr requires a time base larger than the $7$ yr of the present
one. For the following analysis we define the SVP of each QSO  as the
mean value $\beta_m$ in a single bin $0 < \tau < 1000$ d. 
Since $\beta(\tau_{ij})$ values are not independent, the  above computation
of the uncertainty represents an overestimate of the standard deviation of 
$\beta_m$.
In Figure 6 $\beta_m$ of the 42 QSOs are reported versus the relevant time
average $\bar{\alpha}$.
The curves are described in the following paragraphs, except the dot-dashed
line representing a black body, which is reported for comparison with the 
results of \cite{tre01}. In the case of a black body (of fixed area), 
defining $x \equiv h\nu/kT$, with $T$, $h$, $k$ equal to temperature, Planck 
and Boltzmann  constants respectively, the spectral slope is 
$\alpha_{BB}(x)= 3-xe^x/(e^x-1)$ and the SVP is
$(d\alpha/dT)/(d\log B_{\nu}/dT)=(\ln 10)[1-x /(e^x-1)]\equiv\beta_{BB}(x)$.
We stress that the dotted line, defined by the above expression,
does not represent a fit to the points (no free parameters). For a fixed
frequency $\log \bar{\nu} \equiv \log (\nu_B \nu_R)^{\frac{1}{2}}$ 
and increasing temperature, a point
moves on the curve from 
left to right (Rayleigh-Jeans limit $\alpha=2, \beta=0$). The ``average QSO'', 
$\langle\bar{\alpha}\rangle = -0.2\pm 1.0, \langle\beta_m\rangle = 2.2\pm 0.9$,
can be represented by temperature changes of a black body of 
$T\approx 10^4$ K, in approximate agreement with the result of \cite{tre01},
which was obtained with different QSO sample and mean rest-frame frequency.

\section{THE EFFECT OF THE HOST GALAXY}

An independent analysis of the same light curves, performed by \cite{cid00} 
in the framework of Poissonian model of variability, implies the existence of
an underlying spectral component, redder than the variable one, which could
be identified either with the non-flaring part of the QSO spectrum or with 
the host galaxy considered by \cite{rom98}.
We evaluate the effect of the host galaxy through numerical simulations
based on templates of the QSO and host galaxy SEDs \citep{tre01b}.
Both SEDs are derived from the atlas of normal QSO continuum spectra
of \cite{elv94}. We compute a synthetic QSO+host spectrum, adding 
to the fixed host galaxy template SED the average QSO spectrum with a
relative weight characterized by the parameter 
$\eta \equiv  \log({L_H^Q}/{L_H^g})$,
where ${L_H^Q}$ and ${L_H^g}$ are the total $H$ band luminosities of the  
QSO and the host galaxy respectively.   In the \cite{elv94} sample,
$\eta$ ranges from -1 to 2. Figure 5 shows an example of composite spectrum.
We want to test (disprove) the hypothesis that the QSO SED maintains its
shape during brightness changes, while the variation of the spectral shape
is entirely due to the contribution of the (constant) galaxy SED.
We know that the effect of the host galaxy should be small in the case
of \cite{giv99} data since: i) magnitudes were computed using point spread 
function fitting of the images, ii) the images were limited by an aperture
with the diameter depending on the seeing conditions, but smaller than 3 
arcsec. However we don't know the appropriate value of $\eta$.
For this reason we perform  simulations for a range of $\eta$ values.
Variability is represented by small changes $\Delta \eta$, around each
$\eta$ value, with an amplitude corresponding
to a r.m.s. variability $\sigma_B = 0.16$ mag in the blue band. 
For each synthetic spectrum, representing the QSO plus host SED at a given
time, we compute $\alpha(\bar{\nu},t) \equiv
(\partial \log L_{\nu}/\partial\log\nu)_{\nu=\bar{\nu}}$,
$\bar {\nu} = \sqrt {\nu_B \nu_R}$, then we derive the SVP $\beta$.
The result, for $-3 < \eta < 3$, is shown in Figure 6.
To check the dependence of the result on QSO redshift, the same computation
is repeated for the maximum redshift of the \cite{giv99} sample, $z=0.4$, 
and shown in the same figure.
Although an appropriate choice of $\eta$ can reproduce the observed $\beta$ 
or $\alpha$ separately,
the curves are clearly shifted respect to the distribution of the
observational points.
This means that the effect of the host galaxy is not sufficient to account for
the observed changes of the spectral shape. Thus the spectral variability is 
intrinsic of the active nucleus. 
This also implies that the constant red continuum, resulting from the analysis
of \cite{cid00}, cannot be identified with the host galaxy.
It must be, at least in part, due to the nucleus, 
and can be identified with the spectrum of the non-flaring part of the 
accretion disk.

\section{CHANGES OF $\dot{M}$ AND ``HOT SPOTS''}

Once the spectral variability is ascribed to the nuclear component, we can ask
whether a change of any  of the  parameters defining an emission model can 
account for the observed variations of the spectral shape .
We considered the  accretion disk model of \cite{sie95}, \cite{fio95},
corresponding to a Kerr metric and modified black body 
SED, which depend on the black hole mass $M$, the accretion rate $\dot{M}$ and
the inclination $\theta$ ($\theta=0 \rightarrow$ face-on).
A grid of models has been considered for $\log M/M_{\odot}=7.0,8.0,9.0,10.0$,
$\dot{m} \equiv \dot{M} c^2/L_E=0.1, 0.3, 0.8$ 
(where $L_E$ is the Eddington luminosity
$L_E= \frac{4 \pi G c m_p}{\sigma_e M}$ with the usual meaning of symbols)
and $\mu \equiv \cos \theta = 1, 0.75, 0.5, 0.25, 0.1$.
A change of $\dot{M}$ produces a variation of both  
luminosity and the SED shape. On purely theoretical grounds, we know that 
the time required for the accretion disk to reach a new equilibrium condition 
with a different $\dot{M}$ value is at least of the order of the sound crossing
time  $t_{sound}\approx 10^3-10^5$ d \citep{cou91}, 
i.e. much longer than typical
variability time scales. Still, it is interesting to see how the spectral
changes between two $\dot{M}$ states compare with the observed ones, 
as done by \cite{tri94}, and by \cite{sie95}. In Figure 6 the two curves on the
bottom right represent $\beta$ versus $\alpha$
for varying $M$ (from right to left) as computed for two different values 
of $\dot{m}$ and for $\mu=1$, from the Kerr metric, modified black body model
of \cite{sie95} (their table 4). The spectral variations are clearly smaller,
on average, than the observed ones. This means that a transition e.g. from
a lower to a higher $\dot{M}$ regime implies a larger luminosity change
for a given slope variation. Notice that a  black body of varying $T$
and fixed area provides larger $\beta$ values consistent with
the observations. Thus, $\beta$ values computed for an increase of
$\dot{m}$ would better compare with the case of  a transition to a hotter
disk with larger area.
Observed slope variations of two objects, NGC 5548 and NGC 3783, have been 
compared with the predictions of an accretion disk model by \cite{tri94}, in 
the UV range. They conclude that the observed points lie roughly on curves 
of constant black hole mass, giving confidence in the accretion disk models.
However we notice that also in this case the distribution of observed points
is steeper than the iso-mass lines, specially in the case of NGC 3783.
Our result on the statistical sample of 42 objects indicates that this is 
indeed a systematic effect.

This result suggests that transient phenomena, like hot spots produced
on the accretion disk by instability phenomena \citep{kaw98},
instead of a transition to a new equilibrium state,
may better explain the relatively large changes of the local spectral slope,
The available models of instability phenomena do not provide a spectrum 
of the hot spot. Thus we try a simple ``model'' based on the  addition
of a black body flare to the disk SED, represented by the average 
QSO SED of \cite{elv94} (shown in Figure 5).
The free parameters are temperature $T_{BB}$, and emitting area $A$,
while the constraints are the amplitude of the luminosity change (e.g. in
the $B$ band) and the relevant $\beta$ value (or the relevant $\Delta \alpha$).
The solution is not univocal, given the spread of the observed $\beta$ values.
However, $\Delta B = 0.16$ mag (corresponding to the r.m.s. variability of 
the sample) can be obtained by a  hot spot of $T_{BB} \approx 2\cdot 10^5$ K 
and $A=5 \cdot 10^{30}$ cm$^2$, producing $\beta= 3.2$, or
$T_{BB} \approx 2\cdot 10^4$ K, $A=1,3 \cdot 10^{32}$ cm$^2$, giving
$\beta= 2.2$. This is shown by the large filled squares in Figure 6.
A sudden heating of a fraction of the disk surface is thus capable of 
producing the observed change of the SED in the $B$ and $R$ bands and the 
relevant intra-QSO $\alpha$-L correlation.

\section{SUMMARY AND CONCLUSIONS}

We have performed a new analysis of the $B$ and $R$ light curves of a sample
of 42 PG QSOs made available by the Wise Observatory group \citep{giv99}.
We have shown the existence of a positive v-z correlation when the analysis 
is restricted to a small luminosity bin.

This correlation disappears in the sample as a whole
due to the interplay of v-L, L-z and v-z correlations and the small redhift 
range. The slope of the v-z relation is consistent with the previous 
findings of \cite{dic96}
and \cite{tre01}, thus confirming that the v-z correlation can be entirely
explained by the increase of variability with  frequency, coupled
with the increase of (rest-frame) observing frequency for higher redshift 
objects. The dependence of variability on redshift has been questioned in 
the past (see table 1 of \cite{giv99} for a summary of previous variability 
studies). We stress that: i) it is a relatively weak effect 
($\partial v / \partial z \approx 0.1$) so that it cannot be detected
unless other competing effects (including v-L and z-L correlations) are 
properly taken into account; ii) to get rid of spurious effects connected
with cosmological time dilation and finite sampling time, variability must
be quantified by an intrinsic index defined on the basis of the rest frame 
structure function. Once these prescriptions are adopted, the v-z correlations
as measured in different optical-UV bands and different QSO samples appear 
consistent. This was already found by \cite{dic96} and \cite{tre01}. The
present evidence, though marginal,  is again quantitatively consistent and 
confirms  previous results.

The two-color light curves of the \cite{giv99} sample allow for the first 
time the statistical analysis of SED variability of individual QSOs 
and the study of the spectral slope changes among QSOs of different luminosity,
leading to the evidence of an intra-QSO and an inter-QSO correlation.

We have analyzed the spectral variability by the distribution of the  
parameter $\beta(\tau)\equiv 
\frac{\alpha(t+\tau)-\alpha(t)}{\log L_B(t+\tau)-\log L_B(t)}$
as a function of the spectral slope $\alpha$.
We have compared with the observed distribution the spectral slope changes
produced by the contribution of the host galaxy to the QSO SED, under the 
assumption that
the nuclear spectrum maintains its shape while changing its brightness.
We conclude that the host galaxy alone cannot be responsible for the observed 
spectral changes. Thus the spectral variation must be intrinsic of the
nuclear component.
The $\beta$-$\alpha$ distribution has been also compared with the SED changes 
of a disk model for $\dot{M}$ variations. The latter appear insufficient to 
explain the observed spectral changes.
Bright spots on the disk, likely produced by local instabilities, are able to 
represent the observed spectral variability.
 
Since it is likely that different physical phenomena are causing variability 
in different bands and time scales, multi-frequency analyses will be 
ultimately needed to obtain a complete description. However even a two 
optical bands analysis,
once performed on a statistical sample, provides  valuable constraints 
on the origin of variability. This  strongly suggests to extend the work of 
the Wise Observatory group both in frequency and sampling time
to allow a more detailed comparison with possible models.
 
\acknowledgments
We are grateful to the Wise Observatory Group for promptly making available 
their data to the community, and for providing us with details about the 
observations. 
We are indebted to Fabrizio Fiore for his help in the use of disk models and 
for clarifying discussions.
This research has made use of the NASA/IPAC Extragalactic Database (NED) 
which is operated by the Jet Propulsion Laboratory, California Institute of 
Technology, under contract with the National Aeronautics and Space
Administration.

\clearpage
\figcaption[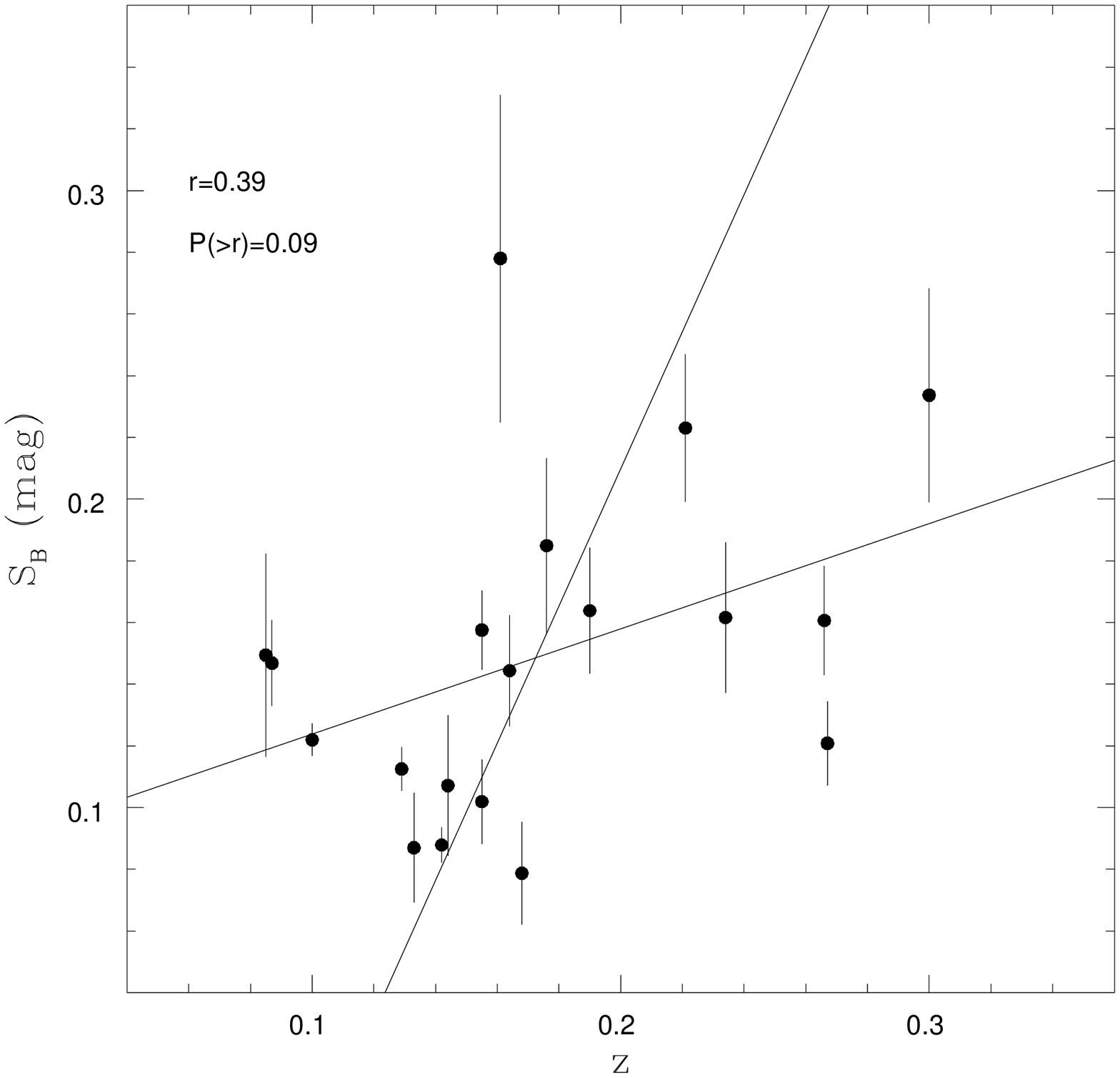]{Variability $S_B$ ( $0.3 \pm 0.09$ yr) versus redshift 
for the subsample with
$-23.5 < M_B < -22.5$. The correlation coefficient $r$ and the relevant 
probability (of the null hypothesis) are also reported.
\label{fig1}}

\figcaption[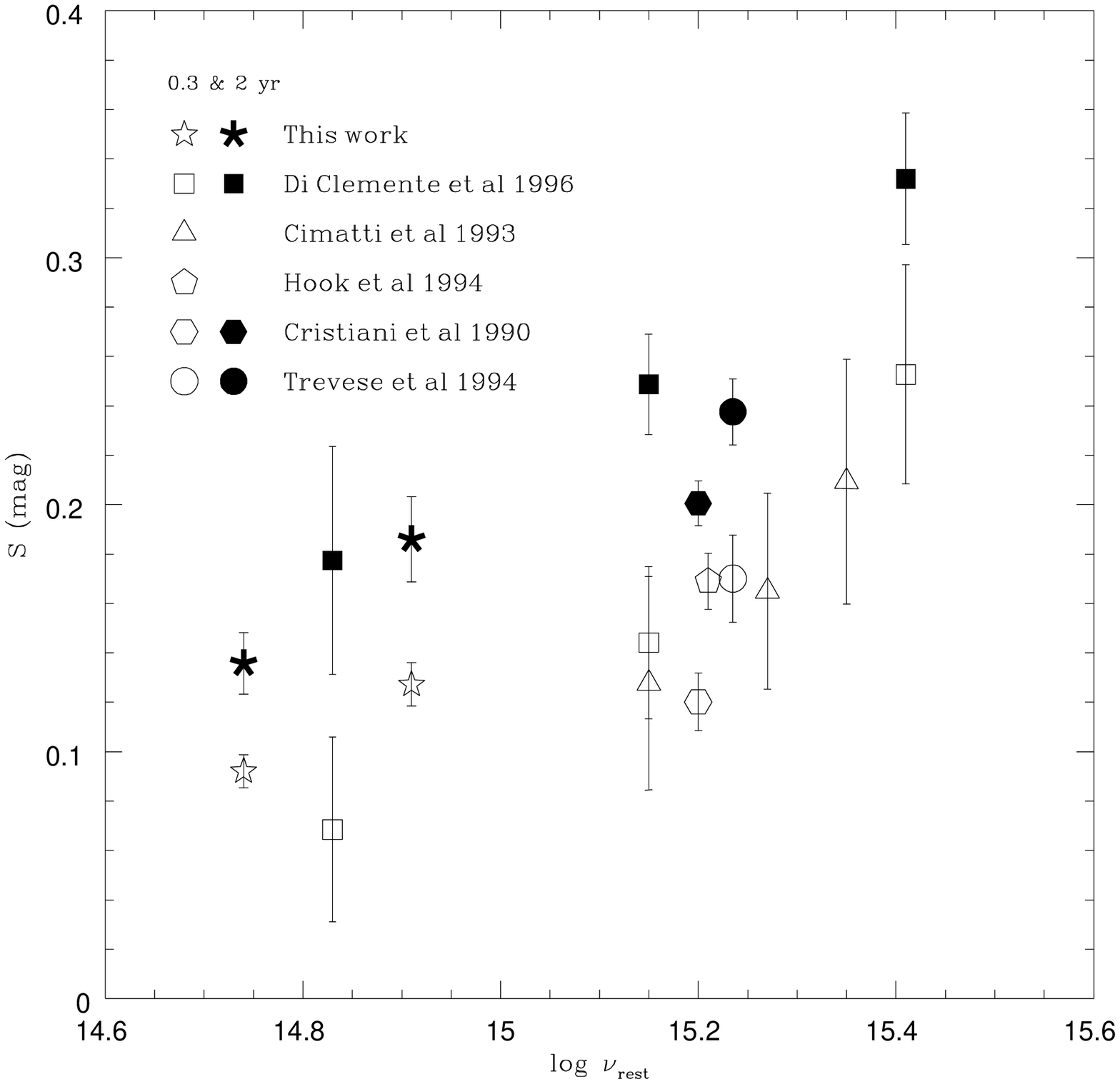]{Variability $S$ versus rest-frame frequency for various
QSO samples (adapted from \cite{dic96}). The variability indicator is the
ensemble
average of variability indexes over the relevant QSO samples. The  frequencies 
of each point are defined as the relevant ensemble averages of the rest-frame
QSO frequencies,  according to the individual  redshift and observing band.
Filled symbols correspond to $\tau=0.3 \pm 0.09$ yr and open symbols 
correspond to $\tau=2.0 \pm 0.3$ yr. The  stars represent the four 
variability indexes of the present analysis. 
\label{fig2}}

\figcaption[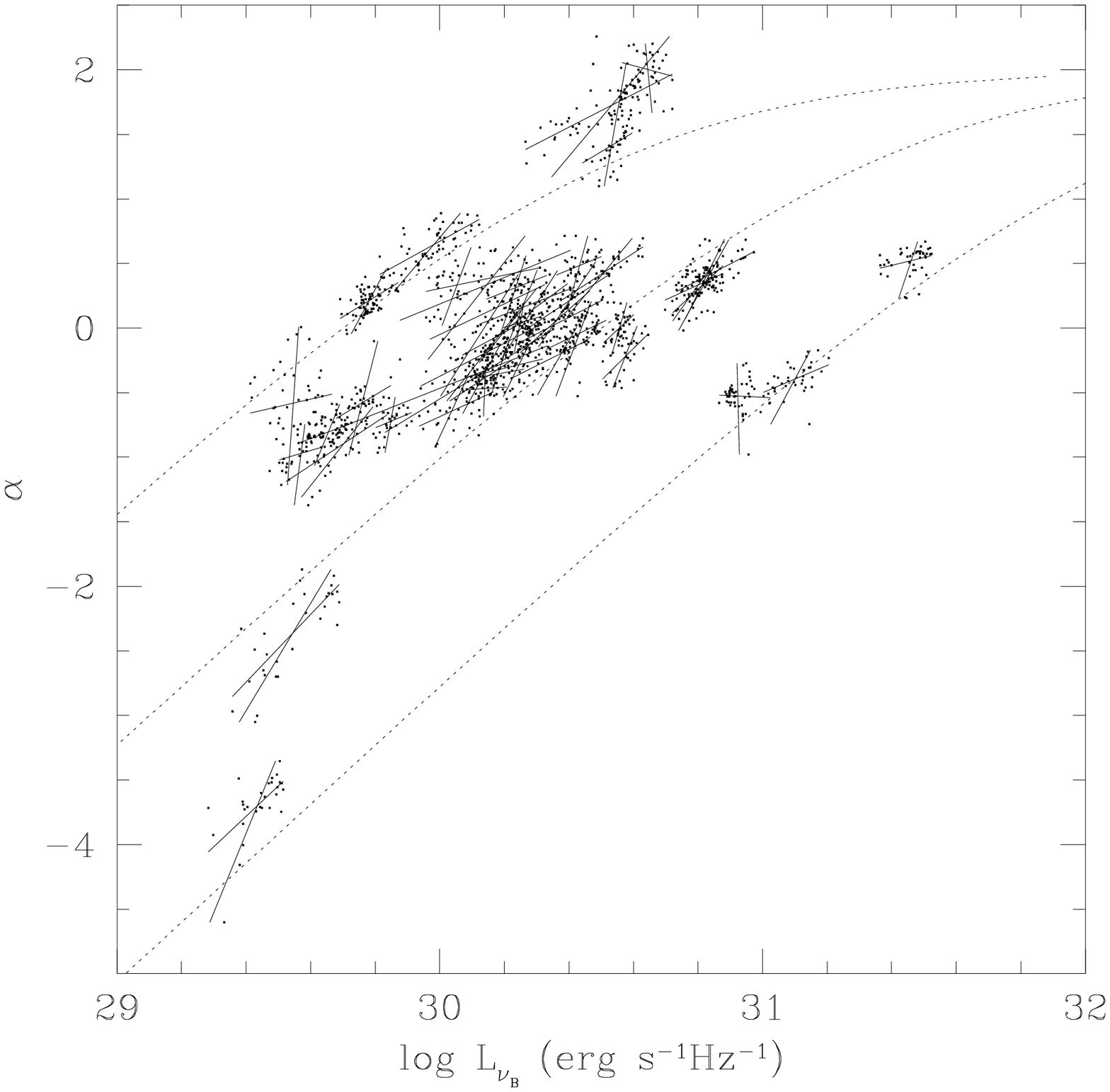]{Instantaneous spectral slope $\alpha$ versus 
monochromatic luminosity $L_{\nu_B}$. Regression lines $\alpha-L_{\nu_B}$ 
are reported for each QSO.
Dotted curves represent black bodies of different areas with temperature $T$ 
increasing from bottom left to top right (Rayleigh-Jeans limit $\alpha=+2$). 
\label{fig3}}

\figcaption[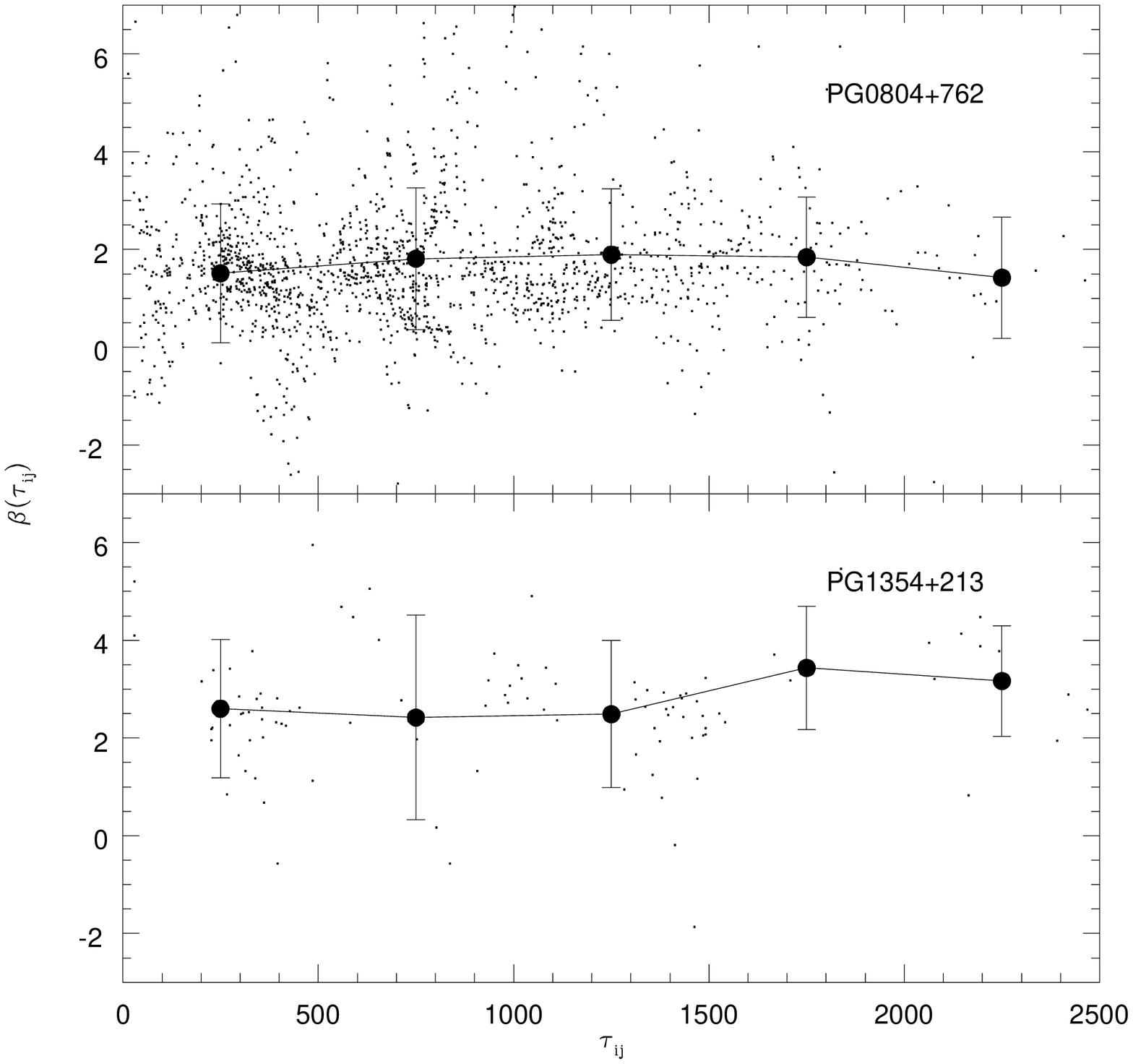]{ The spectral variability parameter $\beta(\tau_{ij})$
as a function of the time lag between the observations for PG0804+762 and 
PG1354+213. 
Filled circles represent the mean value of $\beta(\tau_{ij})$ in intervals 
of 500 d. Error bars represent the r.m.s. uncertainty of the mean.
\label{fig4}}

\figcaption[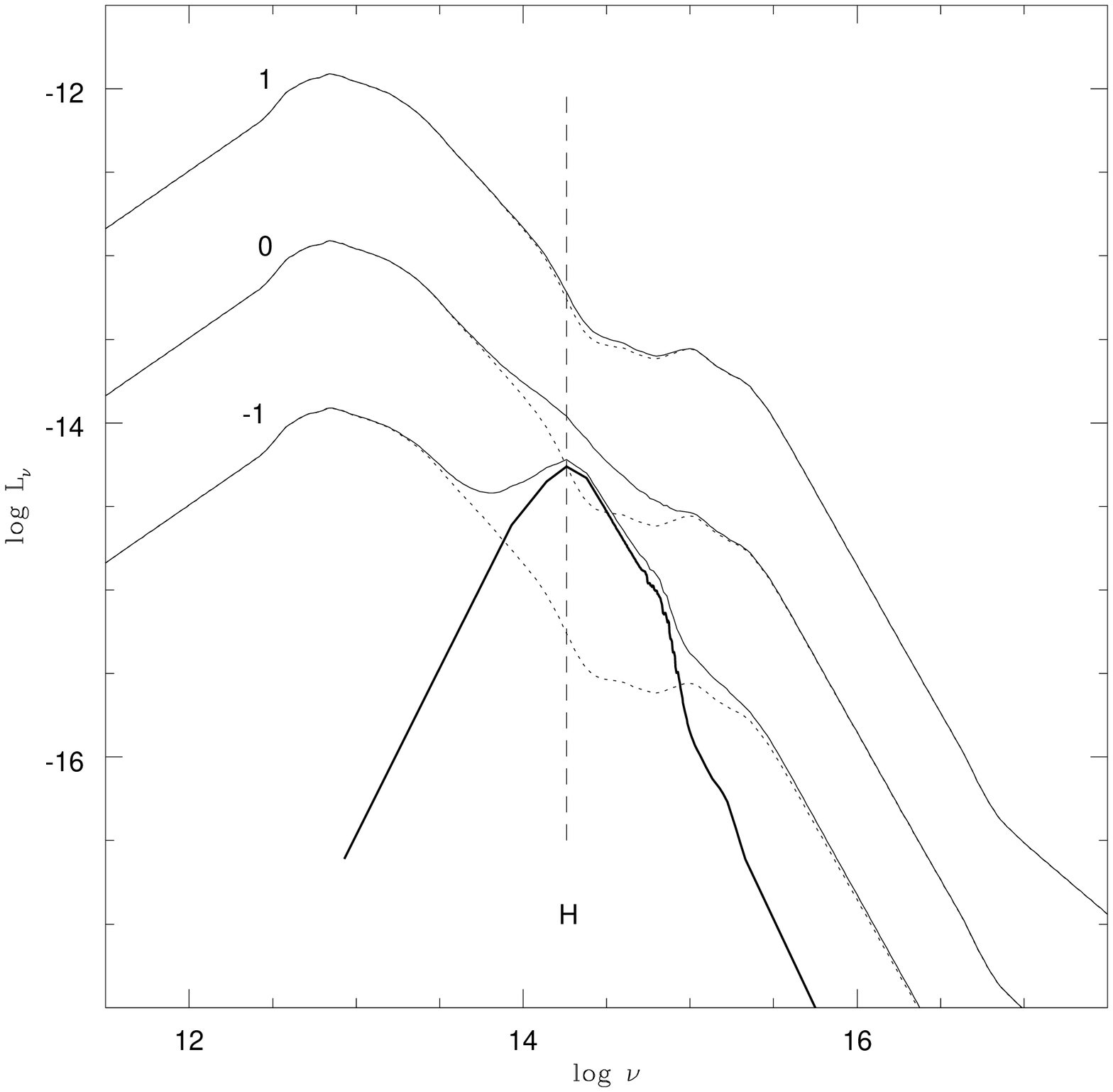]{Synthetic spectra (thin continuous lines) resulting 
from the composition of
the average QSO spectrum  (dotted lines) and the host galaxy template spectrum
(thick line), both taken from
\cite{elv94},  $\eta \equiv  \log({L_H^Q}/{L_H^g}) = -1, 0, 1$
\label{fig5}}

\figcaption[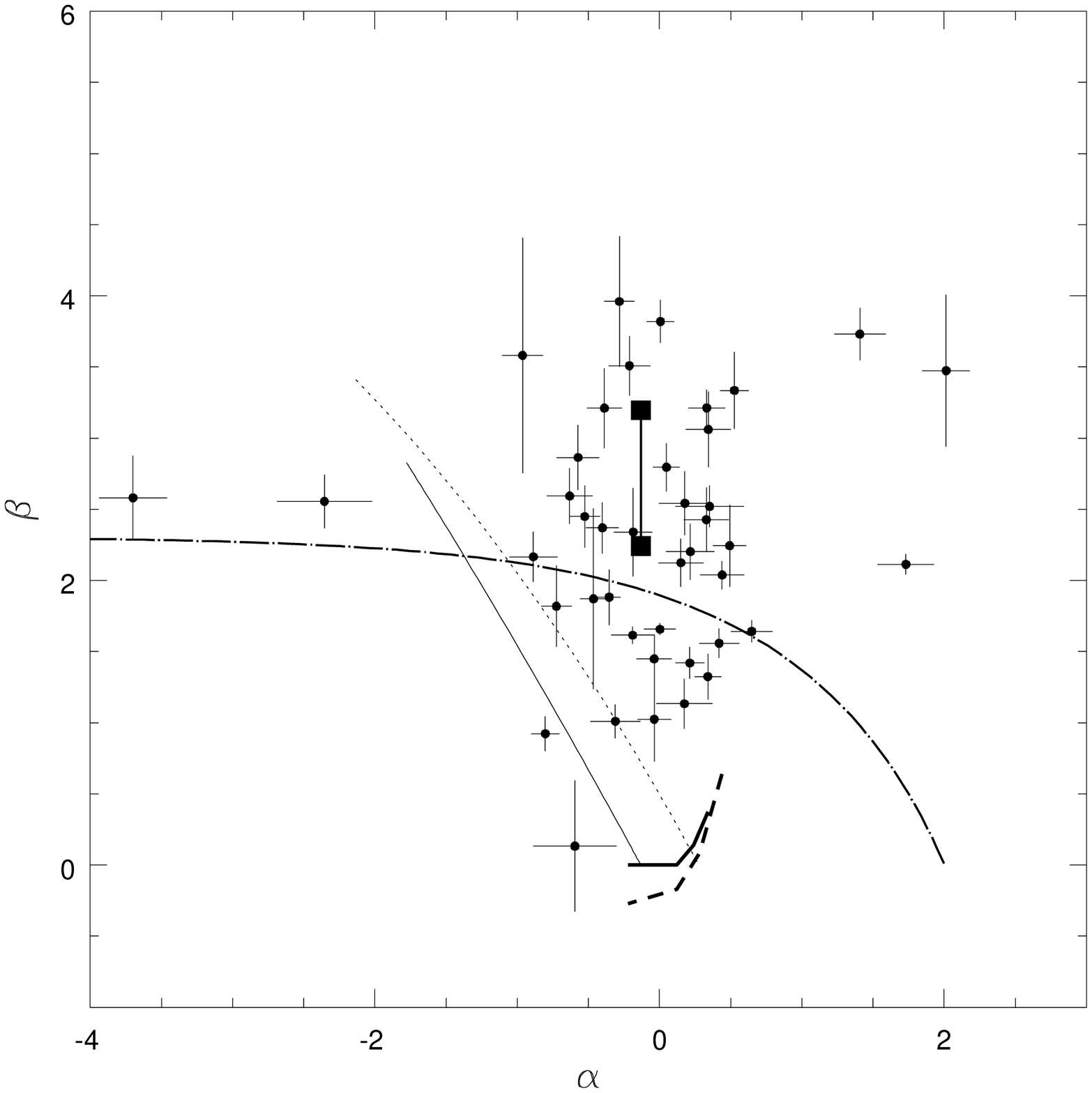]{The spectral variability parameter $\beta_m$ (see Eq. 4) 
versus the average spectral slope for each QSO of the sample. Error bars 
represent the r.m.s.
statistical uncertainties, which are mainly due to intrinsic variability.
Dot-dashed line: sequence of black bodies of different temperatures 
(increasing from left to right);
continuous line: spectral variability due to the presence of the host galaxy, 
in the QSO rest-frame, for  $\eta$  ranging from -3 (left) to 3 (right);
dotted line: the same as continuous line,  for a QSO at the maximum redshift 
of the sample, $z=0.4$;
thick continuous line: spectral variability caused by an increase of   
$\dot{m} \equiv  \frac {\dot{M}}{ \dot{M_{Edd}}}$ from 0.1 to 0.3
in the  model of table 4 of \cite{sie95} (Kerr + modified black body),
$M$ increasing from $10^7$ to $10^{10}$ $M_{\odot}$ (from right to left);
thick dashed line: the same as above for $\dot{m}$ increasing from 0.3 to 0.8.
Large filled squares represent the hot spot model for $T=2 \cdot 10^5$ K 
(upper) and $T=2 \cdot 10^4$ K (lower) (see text).
\label{fig6}}

\clearpage

\begin{figure}[1]
\plotone{f1.eps}
\vskip 0cm
\end{figure}
\clearpage

\begin{figure}[2]
\plotone{f2.eps}
\vskip 0cm
\end{figure}
\clearpage

\begin{figure}[3]
\plotone{f3.eps}
\vskip 0cm
\end{figure}
\clearpage

\begin{figure}[4]
\plotone{f4.eps}
\vskip 0cm
\end{figure}
\clearpage

\begin{figure}[5]
\plotone{f5.eps}
\vskip 0cm
\end{figure}
\clearpage

\begin{figure}[6]
\plotone{f6.eps}
\vskip 0cm
\end{figure}
\clearpage

\end{document}